\documentclass[12pt]{article}
\title{Time-reversal invariance violation measurement using polarized neutron scattering from polarized xenon}
\author{Pinghan Chu\\ 
\small Duke University\\[-0.8ex]}
\date{} 

\def\etal {{\it et~al.}}

\def\PLB{{\rm Phys.~Lett.~} {\bf B}}
\def\PRL{\rm Phys.~Rev.~Lett.~}
\def\PR{\rm Phys.~Rev.~}
\def\PRD{{\rm Phys.~Rev.~} {\bf D}}
\def\PRC{{\rm Phys.~Rev.~} {\bf C}}
\def\PRA{{\rm Phys.~Rev.~}{\bf A}}

\begin{document}

\maketitle

The discoveries of parity ($P$) symmetry violation in nuclear $\beta$ decay~\cite{Wu1957} and charge-parity ($CP$) symmetry violation in $K_0$ decay~\cite{Christenson1964} attracted physicists' attention on time-reversal ($T$) symmetry. If $CPT$ is a good symmetry~\cite{Luders1954}, $CP$ violation implies the existence of $T$ violation. The experiment in $K_0$ decay~\cite{Angelopoulos1998} gave direct evidence of $T$ violation which is consistent with the $CP$ violation parameters in the $K_0$ system. These results can be well explained by the Kobayashi-Maskawa mechanism~\cite{Kobayashi1973} in the Standard Model (SM). However, the contribution of the Kobayashi-Maskawa phase to the $CP$ violation is very small, which fails to fully explain the observed baryon asymmetry of the Universe (BAU)~\cite{Dunkley2009}. Because of Sakharov conditions~\cite{Sakharov1967}, there must be new physics beyond the SM to account for the extra $CP$ and $T$ violation. Therefore, search for the new type of $CP$ and $T$ violation processes becomes of great importance. 

The $CP$ and $T$ violation has only been observed in meson processes; therefore, it is important to test conservation of these symmetries in baryon processes with nucleons. The interaction between polarized neutrons and polarized targets seems to be one of the promising methods to test the $T$ violation. Neutron transmission can be described by the forward neutron-nucleus elastic scattering amplitude~\cite{Stodolsky1986, Beda2007}:
\begin{equation}
f = A + B(\vec{J}\cdot \vec{\sigma}_n) + C(\vec{k}_n\cdot \vec{\sigma}_n)+D\vec{\sigma}_n\cdot(\vec{J}\times\vec{k}_n)+E(\vec{k}_n\cdot\vec{J})+F(\vec{k}_n\cdot \vec{J})(\vec{\sigma}_n\cdot \vec{k}_n\times \vec{J})
\label{eq:neutrontransmission}
\end{equation}
where $\vec{\sigma}_n$ is the neutron spin, $\vec{J}$ is the target spin, and $\vec{k}_n$ is the neutron momentum vector. The coefficient $A$ describes the strong spin-independent interaction. The coefficient $B$ represents the combined effects of the external magnetic field, the spin$-$orbit interaction between the neutron and nucleus, and the interactions of the neutron magnetic moment with nuclear and electron magnetization. The coefficients $C$ and $E$ represent the $P$-violating interaction of polarized neutrons with unpolarized nuclei and unpolarized neutrons with polarized nuclei, respectively. The coefficient $D$ represents the $P$- and $T$-violating interactions. Finally, the coefficient $F$ corresponds to interactions violating $T$ invariance but conserving $P$. 

One advantage of the search for the $T$ violation in nuclei is the possible enhancement of the $T$-violating observables due to complex nuclear structure~\cite{Gudkov1992}. This enhancement arises because of the mixing of strongly excited $s$-wave resonance into a weakly excited $p$-wave. This mechanism is expected to both amplify $P$-violating and $T$-violating effects. Many large $P$-violating effects have been observed in polarized neutron-nucleus scattering for several nuclear species at a certain resonance energy; similar $T$-violating effects are expected to be enhanced at the same resonance energy. 
A $T$-violating test sensitive to the $D$ term in Eq.~\ref{eq:neutrontransmission} requires a polarized nuclear target that possesses a neutron resonance with a large $P$-violating effect for the largest possible enhancement.

Not many nuclei which have large $P$-violating effects are easy to polarize by the required amount. The measurement in~\cite{Szymanski1996, Skoy1996} indicates that an optically-pumping $^{131}$Xe gas target can be used for this experiment. At a resonance energy of $E_n = 3.2$ eV,  the $P$ violation in the n-$^{131}$Xe system is amplified by almost 6 orders of magnitude above its typical amplitude. If the same amplification occurs for a $T$-violating interaction at this resonance energy, the sensitivity to $T$ violation can be comparable to the present limit on the neutron electric dipole moment (EDM).  

The Spallation Neutron Source (SNS) at Oak Ridge National Lab (ORNL) provides the most intense pulsed neutron beams in the world. A polarized neutron beam can be produced by using a polarized $^3$He neutron spin filter, which has been developed at ORNL~\cite{Jiang2013}. The technologies for flipping neutron spin and detecting neutrons already exist. The most difficult challenge of this project is an enriched $^{131}$Xe target with large polarization.

The expression of the $D$ term in Eq.~\ref{eq:neutrontransmission} can be produced by the same mechanisms that produce $CP$ violation in kaon decays. Unlike processes involving elementary particles, however, the interaction in nuclei can be significantly enhanced due to the complex structure of the compound nucleus upon neutron capture. Therefore, the enhancement mechanisms increase the opportunity of observing the effects of symmetry violation.  The effects can be described by a ratio $\lambda_{PT}$ which has the same meaning as the Kobayashi-Maskawa phase $\delta$. The theoretical estimates of $\lambda_{PT}$ are shown in Table~\ref{tab:data} where the prediction of $\lambda_{PT}$ calculated by the Kobayashi-Maskawa phase is about $10^{-10}$  and the upper limits of EDM measurements suggest an upper bound of about $10^{-3}$~\cite{Beda2007}. Because the Higgs boson has been discovered by the LHC at CERN~\cite{CERN2012}, the possibility of discovery of $T$-violating effects in nuclei may lead to new direction in physics.

\begin{table}[t]
		\centering
\begin{tabular}{l|c}
\hline
Model & $\lambda_{PT}$\\
\hline
Kobayashi-Maskawa phase& $\leq 10^{-10}$\\
Left–right symmetry&$\leq 4\times10^{-3}$\\
Horizontal symmetry&$\leq 10^{-5}$\\
Charged Higgs bosons&$\leq 2\times10^{-6}$\\
Neutral Higgs bosons&$\leq 3\times10^{-4}$\\
$\theta$ term in QCD Lagrangian&$\leq 5\times10^{-5}$\\
Neutron EDM (single-loop mechanism)&$\leq 4\times10^{-3}$\\
Atomic EDM (199Hg)&$\leq 2\times10^{-3}$\\
\hline
\end{tabular}
		\caption{Theoretical estimates of $\lambda_{PT}$~\cite{Beda2007}}
\label{tab:data}
\end{table}

The measured asymmetry of $P$ invariance can reach values $\alpha_{P} \sim 10^{-2} - 10^{-1}$ in compound resonances of nuclei~\cite{Alfimenkov1983}, which are many orders of magnitude larger than values $10^{-8} - 10^{-7}$ of parity violation in elementary nucleon-nucleon interactions. In the $p$-wave resonance, the $P$- and $T$- violating asymmetry related to the $D$ term in Eq.~\ref{eq:neutrontransmission} can be written in the form~\cite{Beda2007}
\begin{equation}
\alpha_{PT}\sim\lambda_{PT}\alpha_{P}.
\end{equation}
Using the upper bound $\lambda_{PT}\sim 10^{-3}$ in Table~\ref{tab:data} and $\alpha_{P}\sim 10^{-1}$,  the asymmetry in $P$- and $T$- violation is on the level
\begin{equation}
\alpha_{PT} \leq 10^{-4}.
\label{eq:pt}
\end{equation}
For the $F$ term of $T$-violating with conserving $P$ in Eq.~\ref{eq:neutrontransmission}, the current sensitivity due to the EDM experimental constraint~\cite{Kurylov2001} is about~\cite{Beda2007}
\begin{equation}
\alpha_{T} \leq 10^{-1}.
\label{eq:t}
\end{equation}
Therefore, an experiment with better precision than Eq.~\ref{eq:pt} or Eq.~\ref{eq:t} can compete with EDM experiments. Although the enhancement mechanism was already predicted over twenty years ago, there has been no progress in this field for a long time due to the lack of appropriate techniques~\cite{Beda2007}. With improvements in neutron detection, large intense polarized neutron beams,  and polarized targets in the passed decade, this experiment using polarized neutron scattering from a polarized target has great chance to reveal new physics. 

The first step should be the construction of a $^{131}$Xe gaseous cell and an experiment with unpolarized $^{131}$Xe target similar to~\cite{Szymanski1996, Skoy1996}. The corresponding sensitivity will be estimated and determined based on current technologies. At the same time, research of developing a polarized $^{131}$Xe target should be carried out.  

An enriched $^{131}$Xe target with large polarization is necessary for this experiment. The $^{131}$Xe has a spin $I=3/2$ and possesses a nuclear electric quadrupole moment, which is susceptible to interactions with electric field gradients. Electric field gradients are caused due to the events such as adsorption onto surfaces, distorting the spherical shape of the large electron cloud. Therefore, $^{131}$Xe can serve as a sensitive detector of atomic electron cloud distortions and be used as a surface probe~\cite{Meersmann1998, Pavlovskaya1999}. The quadrupolar interactions dominate $^{131}$Xe longitudinal relaxation, $T_1$, in all phases and cause this relaxation to be significantly faster than that in $^{129}$Xe. Extrapolating from experimental work in~\cite{Brinkmann1962} , a $T_1$ of 25 seconds for pure $^{131}$Xe gas at atmospheric pressure is expected. Hyperpolarized $^{131}$Xe has been produced by spin-exchange optical pumping (SEOP) and studied with optically detected NMR inside of the SEOP cell. In the SEOP cell, quadrupolar splitting of $^{131}$Xe on the order of one Hz has been observed with dependences on cell dimensions~\cite{Wu1987,Wu1988,Wu1990, Raftery1994, Butscher1994}, orientation in the magnetic field~\cite{Wu1988}, chemical treatment of SEOP cell wells~\cite{Raftery1994}, and temperature~\cite{Butscher1994}. So far, hyperpolarized $^{131}$Xe with up to 2.2$\%$ spin polarization can be achieved after separation from the rubidium vapor of the SEOP process~\cite{Stupic2011}. A $^{131}$Xe cell similar to~\cite{Stupic2011} should be the first construction step. The study of polarized $^{131}$Xe can also help us understand the surface effect of relaxation.

\end{document}